\documentstyle[aps,twocolumn,graphicx,ifthen]{revtex}

\newcommand{\myscalebox}[1]{\scalebox{0.4}[0.45]{#1}}

\newcommand{\mysection}[1]{\section{#1}}
\begin{document}
\title{How to evaluate ground-state landscapes of disordered systems
thermodynamical correctly}

\author{Alexander K. Hartmann\\
{\small  hartmann@theorie.physik.uni-goettingen.de}\\
{\small Institut f\"ur theoretische Physik, Bunsenstr. 9}\\
{\small 37073 G\"ottingen, Germany}\\
{\small Tel. +49-551-399570, Fax. +49-551-399631}}

\date{\today}
\maketitle
\begin{abstract}
Ground states of three-dimensional EA Ising spin glasses are calculated
for sizes up to $14^3$ using a combination 
of a genetic algorithm and cluster-exact
approximation. For each realization several independent ground states
are obtained. Then, by applying ballistic search and $T=0$
Monte-Carlo simulations, it is ensured that each ground state appears
with the same probability. Consequently, the results represent the true
$T=0$ thermodynamic behavior. The distribution $P(|q|)$ of overlaps is
evaluated.
For increasing size the width of $P(|q|)$ and the fraction of the
distribution below $q_0\equiv 0.5$ converge to zero. This
indicates that for the infinite system $P(|q|)$ is a delta function,
in contrast to previous results.
Thus, the ground-state behavior is dominated by few large clusters of
similar ground states.
 
{\bf Keywords (PACS-codes)}: Spin glasses and other random models (75.10.Nr), 
Numerical simulation studies (75.40.Mg),
General mathematical systems (02.10.Jf). 
\end{abstract}

\mysection{Introduction}

Recently, a new algorithm 
for studying the ground-state landscape of
finite-dimensional spin glasses \cite{binder86} was introduced \cite{alex2}.
 It could be shown that this method is indeed able to
calculate true ground states \cite{alex-stiff}. The $\pm J$ spin glass
(see below)
exhibits a ground-state degeneracy, i.e. many different ground states
exist for each realization. Results \cite{alex-sg2} describing 
the distribution of the ground states  depend on the statistical 
weights of the states which are determined 
by the algorithm which is used. Usually, different
ground states exhibit different weights
\cite{alex-false}, which is thermodynamically incorrect. 
Here, a new technique is applied which avoids this problem.

In this work, three-dimensional Edwards-Anderson (EA) $\pm J$
spin glasses  are  investigated. They consist of $N$ spins 
$\sigma_i = \pm 1$, described by the Hamiltonian
\begin{equation}
H \equiv - \sum_{\langle i,j\rangle} J_{ij} \sigma_i \sigma_j \quad.
\end{equation}

The sum runs over all pairs of nearest neighbors.
The spins are placed on a three-dimensional (d=3) 
cubic lattice of linear size $L$ with periodic boundary conditions in
all directions.
Systems with quenched disorder of the interactions (bonds)
are considered. Their possible values are $J_{ij}=\pm 1$ with equal
probability. To reduce the fluctuations, a constraint is imposed, so that 
$\sum_{\langle i,j\rangle} J_{ij}=0$.

One of the most important questions is whether many pure states exist
for realistic spin glasses. For the infinitely ranged Sherrington-Kirkpatrik
(SK) Ising spin glass this question was answered positively by the
continuous 
replica-symmetry-breaking mean-field (MF) scheme by Parisi \cite{parisi2}.
But also a complete different model is proposed:
the Droplet Scaling (DS) theory 
\cite{mcmillan,bray,fisher1,fisher2,bovier}
suggests that only two pure states (related by a global flip) exist and
that the most relevant excitations are obtained by reversing large
domains of spins (the droplets). From the ground state point of view 
the existence of many pure states means that
two ground states may differ by an arbitrary number of spins. Otherwise
two ground states would only differ by the spin orientations
in some finite domains,
which is always possible in the $\pm J$ model because of the discrete
structure of the interaction distribution.
A detailed discussion can be found in \cite{newman97}, where the
metastate approach \cite{newman96} is used to thoroughly
 analyze MF,DS and other intermediate scenarios.

While earlier Monte-Carlo (MC) simulations \cite{oldmc} 
suffer from small system sizes or equilibration problems 
\cite{kawashima93},  recent results of simulations
 \cite{marinari96} at temperatures just below $T_c$
 seem to find evidence for the MF picture. 
In \cite{moore98}, by applying a Migdal-Kadanoff approximation, MF
behavior was found for small systems at
temperatures slightly below $T_c$, where the correlation length
exceeds the system size. But by going to lower
temperatures or larger systems 
the DS picture turned out to be more appropriate. 
Consequently, the analysis of true ground states should clarify the issue.
In \cite{berg1} ground states were calculated using multicanonical MC sampling,
but no discrimination between MF and DS could be made because of too small
system sizes. Using cluster-exact approximation true ground states
\cite{alex-stiff} were studied and MF behavior was
found \cite{alex-sg2}. 
But, as mentioned at the beginning,
 these results suffer from the fact that not all ground
states are generated with the same probability \cite{alex-false}.
This would indeed  be the correct sampling method, since all ground-state
configurations have exactly the same energy.

In this work ground states of sizes up to $L=14$ are calculated
and a technique is applied, which guarantees that all ground states
enter the result with the same weight, i.e. the correct $T=0$
thermodynamical behavior is obtained. It will be shown that the main
result changes dramatically: with increasing system size the
ground-state behavior is not explained by the MF scenario.

The method presented here is not only useful when the ground state
calculation is performed using cluster-exact approximation. Also 
other methods like simulated annealing or multicanonical simulation
do not guarantee a priori that each ground state is calculated with the same
probability because always a finite number of steps is
used. Thus, the technique presented here has a wide range of
applications.

The paper is organized as follows: next a short description of
the algorithms is presented. Then the definitions of the 
observables evaluated here are shown. In the main section the results
are presented and finally a summary is given.

\mysection{Algorithms}

The calculation of ground states for three-dimensional spin glasses  
belongs to the  class of the NP-hard problems \cite{barahona82}, 
i.e. only algorithms with exponentially increasing running time
are available. Thus, only small systems can be treated.
The basic method used here 
is the cluster-exact approximation (CEA) technique 
\cite{alex2}, which is a
discrete optimization method designed especially for spin glasses. In
combination with a genetic algorithm \cite{pal96,michal92} this method is
able to calculate true ground states \cite{alex-stiff} up to $L=14$. 
Using this technique one does not
encounter ergodicity problems or critical
slowing down like when using algorithms which are based 
on Monte-Carlo methods. 

But, as mentioned before, by applying pure genetic CEA, one does not obtain
the true thermodynamic distribution of the ground states
\cite{alex-false}, i.e. not all
ground states contribute to physical quantities with the same weight.
For small system sizes up to $L=4$ 
it is possible to avoid the problem by generating all $T=0$ states,
i.e. averages can be performed simply by considering each ground-state
once. Since the ground state degeneracy increases exponentially with the
number $N$ of spins, this is not possible for larger system
sizes. Instead one has to choose a subset of all configurations.
 The following procedure is applied to ensure that 
all ground states appear with the same probability in this selection:

By performing the ballistic-search (BS) algorithm \cite{alex-bs} the
ground states are grouped into {\em clusters}. All states which are
accessible via flipping of free spins, i.e. without changing the
energy,  are considered to be in the same
cluster. It has been shown \cite{alex-bs} that the number of clusters
defined in this way diverges exponentially for the three-dimensional
$\pm J$ spin glass. 
The sizes of these clusters can be estimated quite accurately
using a variant of the BS method \cite{alex-bs} even if only few
ground states per cluster are available. Then a
certain number of ground states is selected from each cluster. 
This number is proportional to
the size of the cluster. It means that each cluster contributes with
its proper weight. The selection is done in a manner that many small
clusters may contribute as a collection as well; e.g. assume that 100 states
are used to represent a cluster consisting of $10^{10}$ ground states, 
then for a
set of 500 clusters of size $10^7$ each 
a total number of 50 states is selected.
This is achieved by sorting the clusters in ascending order. The
generation of states starts with the smallest cluster. For each cluster
the number of states generated is proportional to its size multiplied by a
factor $f$. If the
number of states grows too large, only a certain fraction $f_2$
of the states which have already been selected is kept, the 
factor is recalculated ($f\leftarrow f*f_2$) and the process continues
with the next cluster.

The states representing the clusters are generated by $T=0$ Monte-Carlo
simulation, i.e. iteratively spins are selected randomly and flipped
if they are free. The ground states which have been obtained before
are used as initial configurations for the MC simulation. MC is
able to reproduce the correct thermodynamic distribution, if
the simulation time is long enough. Then, all ground-states within a
cluster are visited with the same frequency. Later it will be shown
that for the
largest size $L=14$ and the largest clusters 100 MC steps per
spin are sufficient. 

Since each cluster appears with
a weight proportional to its size and each ground state within a
cluster appears with the same probability, on total each ground state
 has the same likelihood of being generated. Thus, the correct
 thermodynamic distribution is obtained.

\mysection{Observables}

For a fixed realization $J=\{J_{ij}\}$ of the exchange interactions and two
replicas
$\{\sigma^{\alpha}_i\}, \{\sigma^{\beta}_i\}$, the overlap \cite{parisi2}
is defined as

\begin{equation}
q^{\alpha\beta} \equiv \frac{1}{N} \sum_i \sigma^{\alpha}_i
\sigma^{\beta}_i  \quad .
\label{def_q}
\end{equation}

The ground state of a given realization is characterized by the probability
density $P_J(q)$. Averaging over the realizations $J$, denoted
by $[\,\cdot\,]_{av}$, results in ($Z$ = number of realizations)

\begin{equation}
P(q) \equiv [P_J(q)]_{av} = \frac{1}{Z} \sum_{J} P_J(q) \quad .
\label{def_P_q}
\end{equation}
Because no external field is present the densities are symmetric:
$P_J(q) = P_J(-q)$ and $P(q) = P(-q)$. Therefore, only $P(|q|)$ is relevant.

The Droplet model predicts that only two pure states exist, implying
that $P(|q|)$ converges  to a delta function
$P(q) = \delta(q-q_{EA})$ for $L \to \infty$
(we don't indicate the $L$ dependence by an index), while in the MF picture
the density remains nonzero for a range $0 \le q \le q_1$ with a
peak at $q_{\max}$ ($0< q_{\max} \le q_1$). Consequently the variance

\begin{equation}
\sigma^2(|q|) \equiv \int_{-1}^1 (\overline{|q|} - |q|)^2P(q)dq = 
\overline{|q|^2} - \overline{|q|}^2 \label{def_sigma_q}
\end{equation}
stays finite for $L \to \infty$ in the MF pictures 
while 
$\sigma^2(|q|) \sim L^{-y} \to 0$ according the DS approach. The
combined average of a quantity $X$ 
over all ground states and over the disorder is
denoted with $\overline{X}$. Here, $y$ is the
zero-temperature scaling exponent \cite{mcmillan} denoted $\Theta$ in 
\cite{fisher1,fisher2}.

To characterize the contribution from small overlap values separately, which
are due to a complex structure of the energy landscape, the weight
$X_{q_0}$ of the distribution below a given threshold $q_0$ is calculated:

\begin{equation}
X_{q_0} \equiv \int_0^{q_0} P(|q|) \, dq \quad .
\end{equation}

The overlap defined in (\ref{def_q})
can be applied to measure the distance $d^{\alpha\beta}$ between two states:

\begin{equation}
d^{\alpha\beta} \equiv 0.5(1-q^{\alpha\beta})
\end{equation}
with $0\le d^{\alpha\beta} \le 1$. 
For three replicas $\alpha,\beta,\gamma$ the usual triangular inequality
reads
$d^{\alpha\beta} \le d^{\alpha\gamma} + d^{\gamma\beta}$.
Written in terms of $q$ it reads

\begin{equation}
q^{\alpha\beta} \ge q^{\alpha\gamma} + q^{\gamma\beta}-1 \quad .
\label{triangular_q}
\end{equation}

Another characteristic attributed to the MF scheme is that the state space
exhibits {\em ultrametricity}.
In an ultrametric space \cite{rammal86} 
the triangular inequality is replaced by a stronger one
$d^{\alpha\beta} \le \max(d^{\alpha\gamma}, d^{\gamma\beta})$
or equivalently

\begin{equation}
q^{\alpha\beta} \ge \min(q^{\alpha\gamma}, q^{\gamma\beta}) \quad .
\label{ultra_q}
\end{equation}

An example of an ultrametric space is given by 
the set of leaves of a binary tree:
the distance between two leaves is defined by the number of edges on a path
between the leaves.

Let $q_1\le q_2 \le q_3$ be the overlaps $q^{\alpha\beta}$, 
$q^{\alpha\gamma}$, 
$q^{\gamma\beta}$ ordered according their sizes.
By writing the smallest overlap on the left side in equation (\ref{ultra_q}), 
one realizes that two of the overlaps must be equal and 
the third may be larger or the same: $q_1 = q_2 \le q_3$.
Therefore, for the 
the difference 
\begin{equation}
\delta q\equiv q_2-q_1
\label{def_delta}
\end{equation}
  $\delta q=0$ holds. For a finite system ultrametricity may be
 violated, i.e. $\delta q >0$. If a system becomes more and more
 ultrametric with growing system size, $\delta q$ should
 decrease while $L\to\infty$.
When evaluating $\delta q$, the influence of the absolute size
of the overlaps should be excluded. Thus, the third overlap is fixed: 
$q_3=q_{fix}$. In practice 
overlap triples are used where $q_3 \in [q_{fix},q_{fix2}]$ holds. This
allows to obtain sufficient statistics. In the next section
 the distribution $P(\delta q)$ is evaluated. For
an ultrametric system this quantity should converge  to a 
Dirac delta function with increasing size $L$ \cite{bhatt86}.

\mysection{Results}

Ground states were generated using genetic CEA for sizes 
$L\in[3, \ldots, 14]$. 
The number of realizations of the bonds per lattice size ranged from
100 realizations for $L=14$ up to 1000 realizations
for $L=3$. One $L=14$ run needs typically 540 CPU-min on an 80MHz PPC601
processor (70 CPU-min for $L=12$, $\ldots$, 0.2 CPU-sec for $L=3$),
more details can be found in \cite{alex-stiff}.
Each run resulted in one  configuration which was
stored, if it exhibited the ground state energy. 
For the smallest sizes $L=3,4$ all ground states were calculated for
each realization by performing up to $10^4$ runs.
For larger sizes it is not possible to obtain all ground states,
because of the exponentially rising degeneracy. 
 For $L=5,6,8$ practically all clusters are obtained using at most
$10^4$ runs \cite{alex-bs}, only for about 25\% of the $L=8$
realization some small clusters may have been missed.

For $L>8$ not only the number of states but also the number of clusters
is too large, consequently
$40$ independent runs were made for each realization. 
For $L=14$ this resulted
in an average of $13.8$ states per realization having the lowest energy 
while for $L=10$  on average $35.3$ states were stored. This seems
a rather small number.
However, the probability that genetic CEA returns a specific ground state
increases (sublinearly) with the size of the cluster the state belongs
to \cite{alex-analyse-cea}. 
Thus, ground states from small clusters do appear with a 
small probability. Because the behavior is
dominated by the largest clusters, the results shown later on are the
same (within error bars) 
as if  all ground states were available. This was tested
explicitly for 100 realizations of $L=10$ 
by doubling the number of runs, i.e. increasing the number of clusters found.

Using this initial set of states for each realization ($L> 4$) 
a second set was produced using the techniques explained before, 
which ensures that each ground
state enters the results with the same weight.
The number of states was chosen in a way, that
$n_{\max}=100$ states were available for the largest clusters of each
realization, i.e. a single cluster smaller than one hundredth of the largest
cluster does not contribute to physical quantities, but, as explained
before, a collection of many small clusters contributes to the results as well.
Finally, it was verified that the results did not change by
increasing $n_{\max}$. 

The number of MC steps used for generating the states was
determined in the following way: a ground state was selected randomly
from the largest clusters found for the
$L=14$ realizations. 100 independent
 $T=0$ MC runs of length $n_{MC}$ MC steps were performed starting
always from this initial state. For the set of 100 final states the
distribution of overlaps was calculated. The whole process was
averaged over different realizations. In Fig. \ref{figPqClusterMC}
the average distribution  $P_c(q)$ 
of overlaps is shown for different run lengths
$n_{MC}$. It can be seen that by increasing the number of MC steps
the ground-state cluster is explored better. By going beyond
$n_{MC}=100$ steps $P_c(q)$ does not change, indicating that this
number of MC steps is sufficient to generate ground states  equally
distributed within a cluster for $L=14$.

The order parameter selected here for the description of the complex ground
state behavior of spin glasses is the total distribution $P(|q|)$ of
overlaps. The result for the case where all ground states have the
same weight is shown in Fig. \ref{figPqEqui} for $L=6,10$. The
distributions are dominated by a large peak for $q>0.8$. Additionally
there is a long tail down to $q=0$, which means that arbitrarily
different ground states are possible. So far this is the same result as
obtained earlier \cite{alex-sg2} for the case where the weights of the
states are determined by the genetic CEA algorithm. But there is a
difference: for the old results the weight of the long tail remains
the same for all system sizes. Here for $L=10$ small overlaps are
about $3/4$ times less likely than for $L=6$.

To study the finite size dependence of this effect, the variance 
$\sigma^2(|q|)$ of
$P(|q|)$ was evaluated as a function of the system size $L$. The
result is displayed in Fig. \ref{figSigmaQ}. Additionally the 
datapoints from \cite{alex-sg2} are given. Obviously, by guaranteeing that
every ground state has the same weight, the result changes
dramatically. To extrapolate to $L\to\infty$, a fit of the data to
$\sigma^2_L = \sigma^2_{\infty}  +a_0L^{-a_1}$ was performed. A value
of $\sigma^2_{\infty}=-0.01(1)$ ($a_1=-0.61(15)$) was obtained, 
indicating that the width of $P(|q|)$ is zero for the infinite system.
Consequently, the MF picture with a continuous breaking of replica
symmetry cannot be true for three-dimensional 
$\pm J$ spin glasses.
 
In Fig. \ref{figXLEqui} the behavior of the long tail is studied in
more detail. The integrated weight $X_{0.5}(L)$ of all overlaps
$q<q_0\equiv 0.5$ is
shown as a function of the system size. Again a fit is used to
extrapolate the behavior of the infinite system. 
A value of $X_{\infty}=-0.01(2)$ is obtained, confirming the result
obtained above.

One might suspect that the results can be explained by the fact that
with increasing system size the 
behavior is dominated more and more by one ground-state cluster. To
examine this issue the quantity $Y=1-[ \sum_c w_c^2]_{av}$ 
is calculated, where $w_c$ is the relative size of cluster $c$. If
really one cluster dominates, $Y$ must vanish with increasing system
size $L$. In Fig. \ref{figMeanYL} $Y$ is shown as a function of $L$
for small system sizes $L\le 8$, where all ground-state clusters have
been obtained.
Obviously, $Y$ does not decrease. One reason is that the probability
$P(n_c=1)$ that a realization exhibits just one ground-state cluster (and
its inverse) decreases with growing system size (cf.
inset). Consequently, there is no single reason explaining the
behavior of $P(|q|)$. Additionally, for the interpretation of
Fig.\ref{figMeanYL},  one has to take into account that
the definition of a cluster, although it is very useful
for the evaluation of the ground-state landscape, may have no physical
meaning.

By collecting all results one obtains the following description for the
distribution of overlaps of the infinite system: It consists of a
large delta-peak and a tail down to $q=0$, but the weight of that tail
goes to zero. This expression is used to point out that by going to
larger sizes small overlaps still occur: the number of arbitrarily 
different ground states diverges \cite{alex-bs}. 
But the size of the largest clusters,
which determine the self overlap leading to the large peak, diverges
even faster. The delta-peak is centered around a finite value
$q_{EA}$. From further evaluation of the results $q_{EA}=0.90(1)$ was obtained.

Finally, it was tested whether the ground states are ultrametrically
organized. In Fig. \ref{figPDeltaQ} the distribution  $P(\delta q)$ is
shown for system sizes $L=4,8,12$.  
Each realization enters
the distribution with the same weight. With increasing system
size the distributions get closer to $q=0$, indicating that the
systems become 
more and more ultrametric. The same conclusion can be driven from the
evaluation of the 
average value of $\delta q$ as a function of $L$ (cf. inset).
This result is similar to the former calculations \cite{alex-sg2},
where the correct $T=0$ distribution was not obtained.
 But it should be stressed that ultrametricity is only found
within a restricted subset of states (here $q_3\approx 0.5$). 
By performing
the thermodynamic limit the weight of all regions of state space 
restricted in this way
disappears, i.e. ultrametricity disappears as well.

\mysection{Conclusion}

Using genetic cluster-exact approximation the ground-state landscape
of three-dimensional $\pm J$ spin glasses is investigated. By applying
ballistic search and $T=0$ Monte-Carlo simulation it is guaranteed that
each ground state enters the result with the same probability, thus a
correct thermodynamic distribution is achieved. This technique also
can be successfully combined with other methods which are used to
generate several configurations from  a degenerate ground-state
landscape, e.g. with simulated annealing or
multicanonical simulation.

The distribution of overlaps is evaluated. For the infinite system it
consists solely of two symmetric delta-peaks. This does not imply that
there are only two ground-state clusters remaining. On the contrary,
the number of
ground state clusters grows exponentially with increasing system size  but
the ground-state behavior is dominated by a few large similar clusters
(and their inverse). Therefore, a distinct impression emerges:
a huge number of arbitrarily different
ground-state clusters exist, but by going to larger
and larger sizes most of them become unimportant. This rules out any
(nonstandard) MF picture with continuous breaking of symmetry 
to be valid in total. Interestingly, the result is compatible with the
one step replica-symmetry-breaking scheme which was 
observed for the p-spin glass \cite{gross84}. It exhibits a simple
distribution of overlaps while many different ground-state clusters
are possible. However,
further work is needed to determine which
of the remaining scenarios really holds for finite-dimensional spin glasses.

Please note that the cluster-interpretation  depends on the definition of a
cluster. By choosing a dynamic which allows flips of more than one spin
at a time, a different definition of energy-barriers is implied and thus
another kind of clusters. But
it should be stressed that the results presented in this work do 
  not depend on the way a cluster is defined. Any method of sorting
the ground states into groups will work that takes the number of ground states
selected proportional to the size of the group, and ensures that each state
of a group has the same probability of being used for the calculation.

Finally, it should be pointed out that not all results previously
obtained using genetic CEA are biased by the disbalance of the
ground-state distribution. The main outcomes in
\cite{alex-stiff,alex-threshold} are
not affected. 
Additionally, although the old data bases on a wrong
distribution, the results in \cite{alex-sg2}
prove that there are arbitrary different clusters present. The reason for
$P(|q|)\to \delta(q-q_{EA})$ is that
 most of them become less important.

\mysection{Acknowledgements}

The author thanks T. Aspelmeier, K. Bhattacharya, M. Otto and
 A. Zippelius for interesting discussions.
He is gratefull to A. Zippelius and O. Herbst 
for critically reading the manuscript.
The work was supported by the Graduiertenkolleg
``Modellierung und Wissenschaftliches Rechnen in 
Mathematik und Naturwissenschaften'' at the
{\em In\-ter\-diszi\-pli\-n\"a\-res Zentrum f\"ur Wissenschaftliches
  Rechnen} in Heidelberg and the
{\em Paderborn Center for Parallel Computing}
 by the allocation of computer time. The author obtained financial
 support from the DFG ({\em Deutsche Forschungs Gemeinschaft}) under
grant Zi209/6-1.


\newcommand{\captionPqClusterMC}
{Distribution $P_c(q)$ of overlaps restricted to a ground-state
  cluster for different number $n_{MC}$ of MC steps. 100 independent
 $T=0$ MC runs were performed for the largest
clusters found for $L=14$ starting
always from the same ground state. By going beyond $n_{MC}=100$ the
distribution does not change any more, indicating that 100 MC steps
are sufficient to obtain independent ground states within a
cluster for $L\le 14$. 
The inset shows the mean overlap value as a function of $n_{MC}$. 
}
\newcommand{\captionPqEqui}
{Distribution $P(|q|)$ of overlaps for $L=6,10$. Each ground state
  enters the result with the same probability. The fraction of small
  overlaps decreases about a factor $3/4$ by going from
  $L=6$ to $L=10$ (please note the logarithmic scale).}

\newcommand{\captionSigmaQ}
{Variance $\sigma^2(|q|)$ of the distribution of overlaps as a function
  of linear system size $L$. The upper points show the case were each
  ground state enters with a weight determined by the genetic CEA
  algorithm. For the lower points each ground state has the same
  probability of being included in the calculation. The line
  represents a fit to $\sigma^2_L = \sigma^2_{\infty}
  +a_0L^{-a_1}$. The extrapolation to the infinite system results in 
$\sigma^2_{\infty}=-0.01(1)$. Consequently, the width of distribution of
overlaps appears to be zero, i.e. $P(|q|)$ is a delta-function.}

\newcommand{\captionXLEqui}
{Fraction of the distribution of overlaps below $q_0\equiv 0.5$ as a
  function of linear system size $L$. Here the results for the correct
  thermodynamic average is shown. The line
  represents a fit to $X_{0.5}(L) = X_{\infty}
  +x_0L^{x_1}$. The extrapolation $L\to\infty$ results in 
$X_{\infty}=-0.01(2)$, i.e. for the infinite systems small
values of $|q|$ occur with frequency zero.}

\newcommand{\captionMeanYL}
{Cumulant $Y$ describing the distribution of cluster sizes as a
  function of $L$. For small system sizes $L\le 8$ almost 
  all ground-state clusters
  have been obtained. The figure proves that with increasing system
  size the ground-state landscape is not dominated more and more by
  one cluster. The inset shows the probability $P(n_c=1)$ that a realization
  exhibits only one ground-state cluster as a function of L.
}

\newcommand{\captionPDeltaQ}
{Distribution $P(\delta q)$ for different system sizes $L=4,8,12$ 
where $\delta q=q_2-q_1$ and $q_1\le q_2\le q_3$ are
triplets of absolute values of overlaps from 
independent triplets of ground states.
Only triplets with $q_3 \in [0.5,0.6]$ are used. For an infinite ultrametric
system $\delta q=0$ holds. For $L=12$ a running average was
used to smooth the data to make the figure more readable.

With increasing system size
the distributions get closer to $q=0$ indicating the increasing 
ultrametricity
of the ground states. The lines are guides for the 
eyes only. The inset shows the 
average value of $\delta q$ as function of system size $L$. 
The straight line represents the
function $\langle \delta q \rangle(L)=0.229 \times L^{-0.24}$.
}

\begin{figure}[htb]
\begin{center}
\myscalebox{\includegraphics{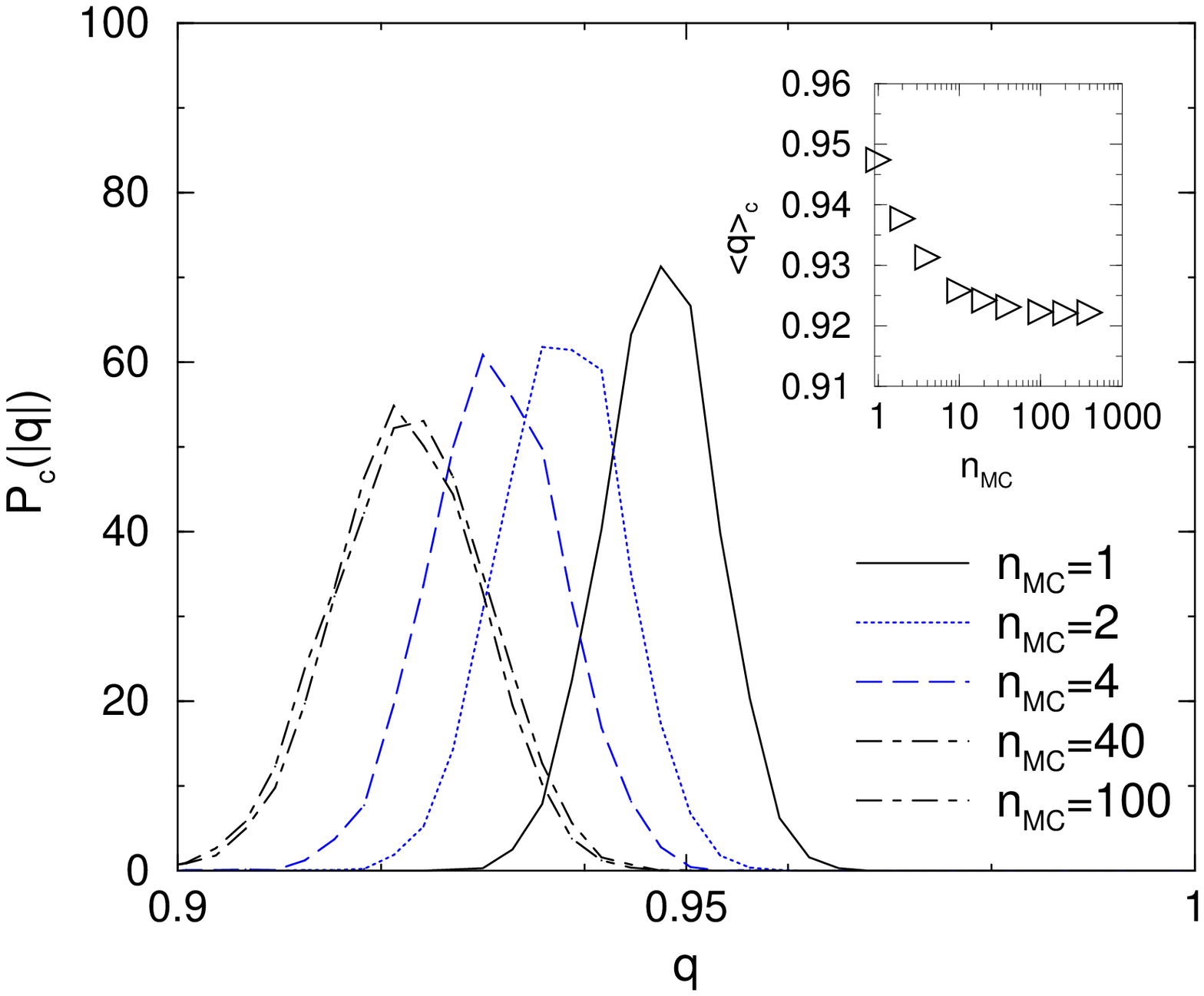}}
\end{center}
\caption{\captionPqClusterMC}
\label{figPqClusterMC}
\end{figure}

\begin{figure}[htb]
\begin{center}
\myscalebox{\includegraphics{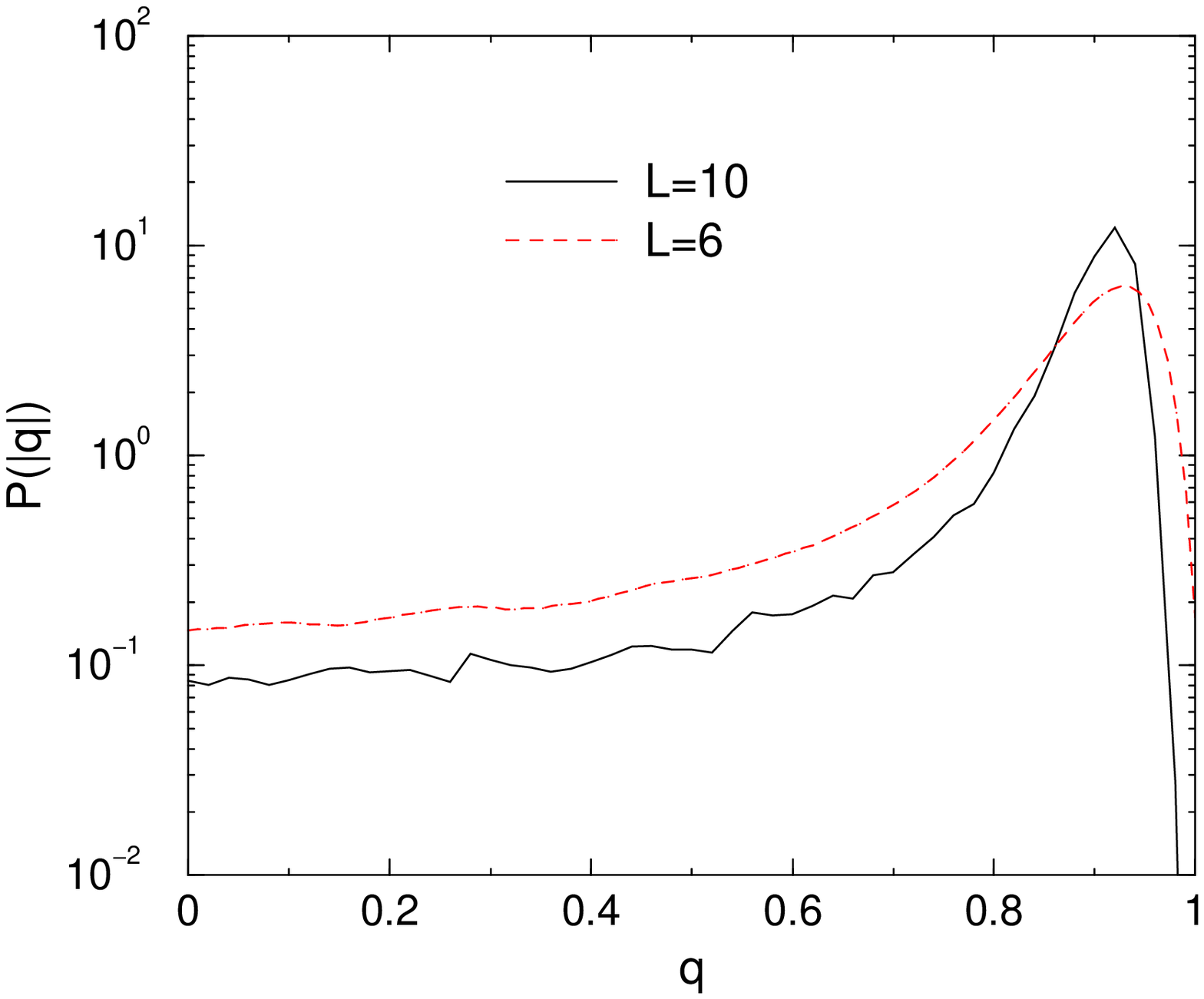}}
\end{center}
\caption{\captionPqEqui}
\label{figPqEqui}
\end{figure}

\begin{figure}[htb]
\begin{center}
\myscalebox{\includegraphics{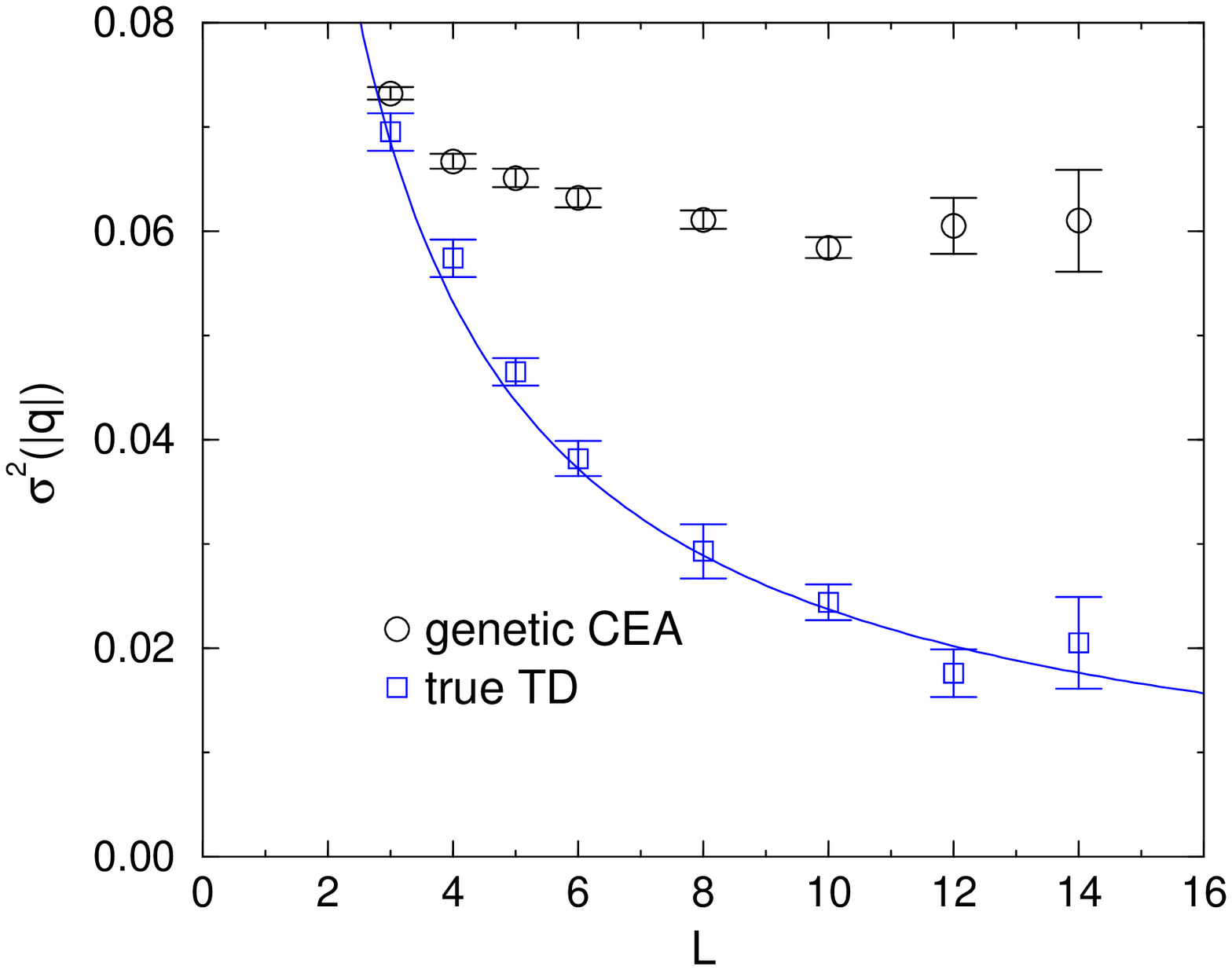}}
\end{center}
\caption{\captionSigmaQ}
\label{figSigmaQ}
\end{figure}

\begin{figure}[htb]
\begin{center}
\myscalebox{\includegraphics{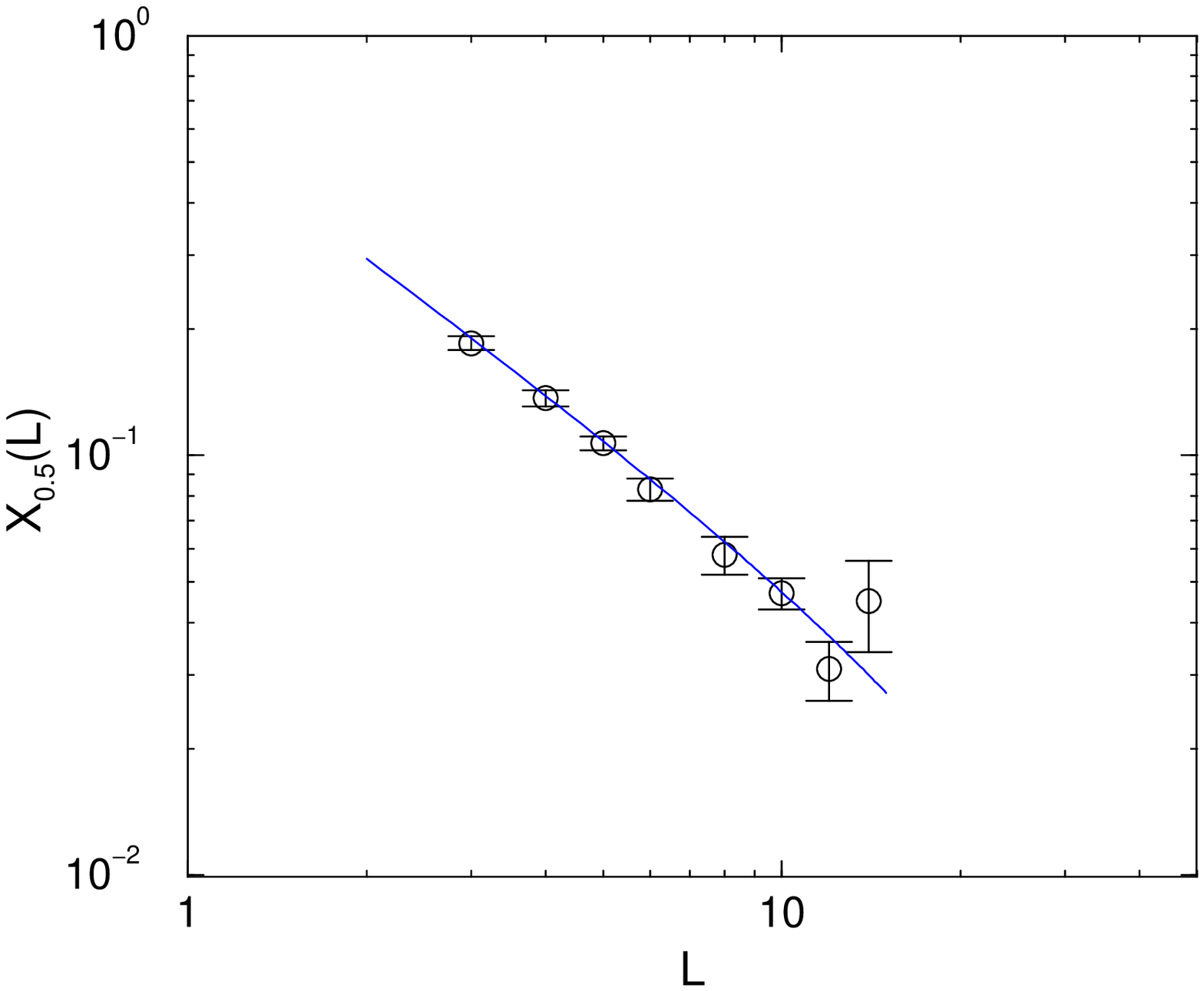}}
\end{center}
\caption{\captionXLEqui}
\label{figXLEqui}
\end{figure}

\begin{figure}[htb]
\begin{center}
\myscalebox{\includegraphics{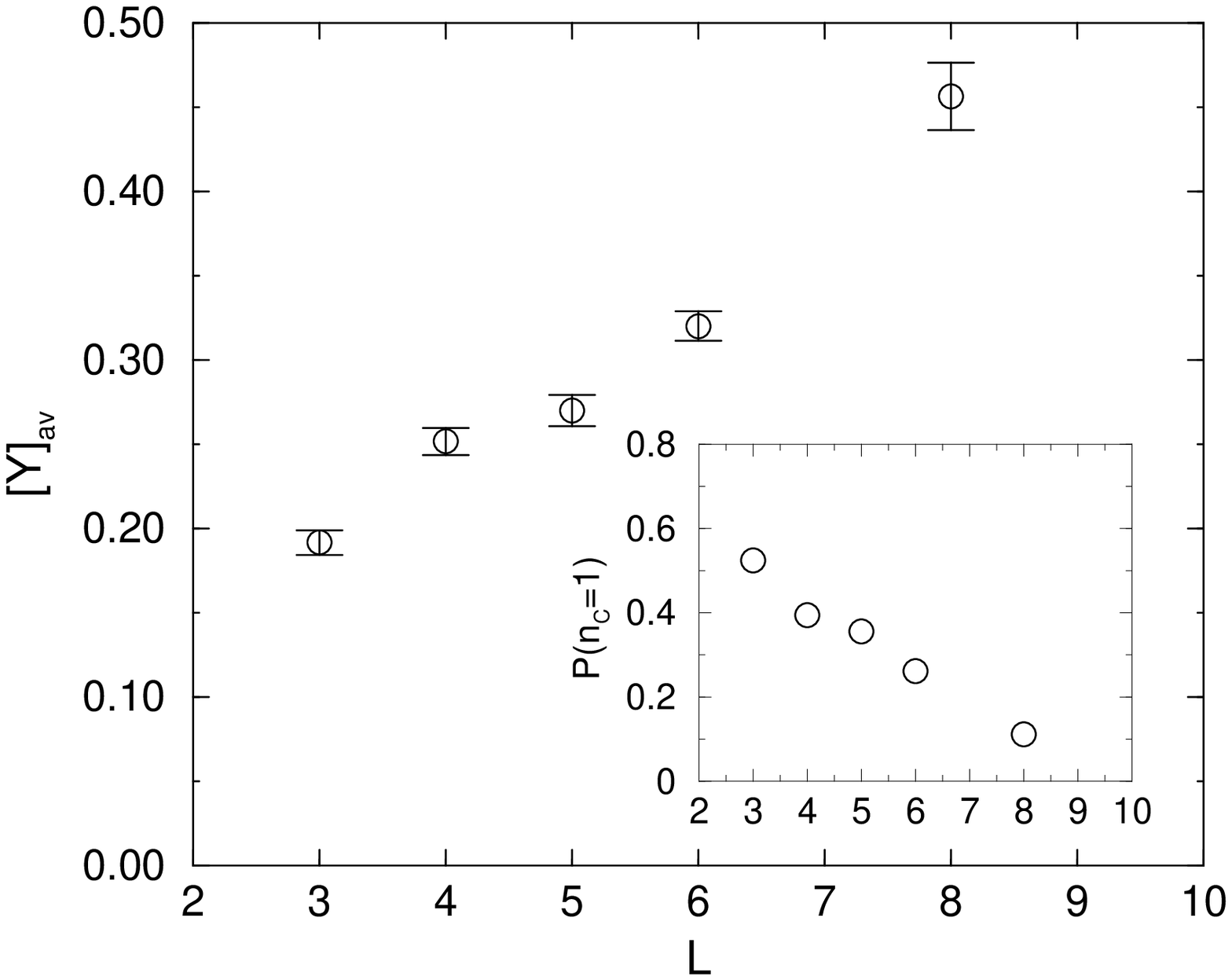}}
\end{center}
\caption{\captionMeanYL}
\label{figMeanYL}
\end{figure}

\begin{figure}[htb]
\begin{center}
\myscalebox{\includegraphics{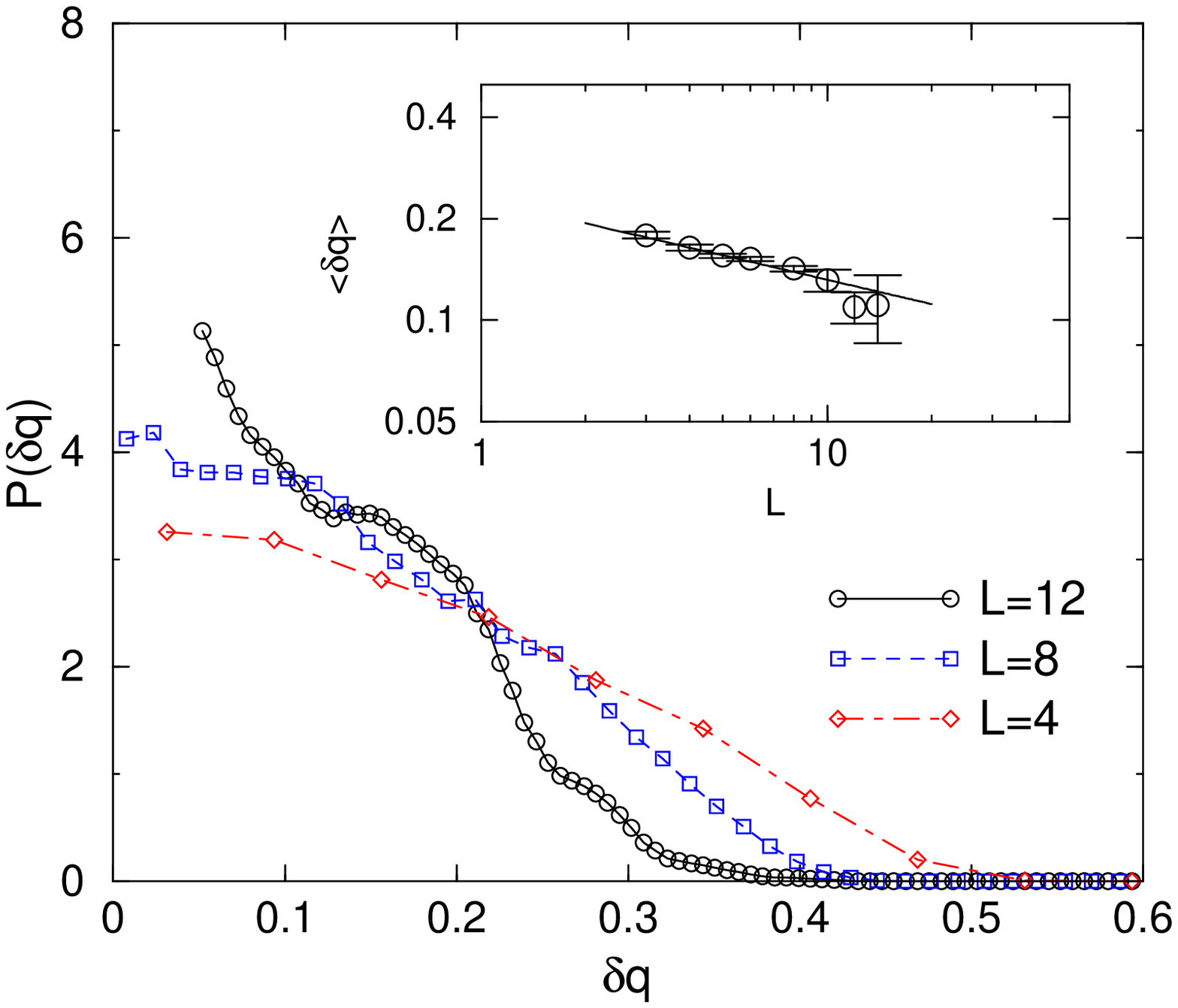}}
\end{center}
\caption{\captionPDeltaQ}
\label{figPDeltaQ}
\end{figure}

\end{document}